\documentstyle[12pt,twoside]{article}
\normalbaselines
\font\tbf = cmbx12

\begin{document}

\indent
\vskip 1cm
\centerline{\tbf SINGULAR BEHAVIOUR OF ELECTRIC AND MAGNETIC FIELDS}
\vskip 0.3cm
\centerline{\tbf IN DIELECTRIC MEDIA}
\vskip 0.3cm
\centerline{\tbf IN  A NON-LINEAR GRAVITATIONAL WAVE BACKGROUND}
\vskip 0.8cm
\centerline{by}{}
\vskip 0.3cm
\centerline{\tbf Alexander B. Balakin\footnote{e-mail: dulkyn@mail.ru} }
\vskip 0.3cm
\centerline{\it Kazan State University, Kremlevskaya street 18, 
420008, Kazan,  Russia,}
\vskip 0.5cm
\centerline{and}
\vskip 0.5cm
\centerline{\tbf Jos\'e P. S. Lemos\footnote{e-mail: lemos@kelvin.ist.utl.pt} }
\vskip 0.3cm
\centerline{\it CENTRA, Departamento de F\'{\i}sica, Instituto Superior 
T\'ecnico,}
\centerline{\it Av. Rovisco Pais 1, 1049-001 Lisboa, Portugal,}
\centerline{\it and}
\centerline{\it Observat\'orio Nacional - MCT,}
\centerline{\it Rua General Jos\'e Cristino 77, 20921, 
Rio de Janeiro, Brazil.}
\vskip 3cm
{\tbf Abstract} 
\indent 
{\small Evolution of  electric and magnetic fields in  
dielectric media, driven by the influence of a strong  
gravitational wave, is considered for four exactly  
integrable models. It is shown that the gravitational wave field 
gives rise to new effects and to singular behaviour 
in the electromagnetic field.}
\vskip 7.5cm

\newpage

\section{Introduction}

In 1970 Bocaletti et al opened up the discussion about the interaction
of weak gravitational wave (GW) fields and static electromagnetic
fields in vacuum \cite{boc}.  In the seventies this problem has been
studied in detail in the application to GW detection (see, e.g.,
\cite{zel1}).  The idea of these investigations was based on two
suggestions (i) that the GW field is weak and (ii) the electromagnetic
field is in vacuum. Item (ii) was motivated by the supposition that
either the presence of matter or vacuum fluctuations make the
velocities of both gravitational and electromagnetic waves different,
therefore destroying the coherence between both types of waves
\cite{zel2} (see, also, \cite{grishpoln}).  

This problem has been taken up in \cite{bala} where items (i) and (ii)
were dropped down and solutions in the presence of matter were
considered for non-linear GW background.  Specifically, an exact
solvable model of evolution of an electromagnetic field in a GW
background in the presence of a spatially isotropic dielectric medium
was studied along with the critical character of the GW effect on the
amplification of the electrodynamical response, (here we are using the
word ``critical'' in analogy to phase transitions terminology).  It
was also shown that one can consider this sort of models in the
framework of nonlinear GW fields \cite{bala}.

One can now go a step further and study the critical behaviour of an
electrodynamical system in a GW background for a wider class of models
other than those considered in \cite{bala}.  We discuss in a covariant
formalism a generalized problem of the singular behaviour of
electrodynamical systems evolving in a GW background.  The paper is
composed as follows. In section 2 the phenomenological representation
of the master equations for electrodynamics of continuous media is
given for arbitrary gravitational background fields and then
considered for four particular models, namely, pure vacuum, curvature
induced corrections in vacuum, spatially isotropic medium, and
curvature induced corrections in spatially isotropic medium.  In
section 3 exact solutions of Maxwell equations in a GW background
(specifically, a pp-wave background) are obtained and discussed for
the four mentioned examples. Attention is given to the
singular behaviour problem. In section 4 we present our conclusions.

\section{Master equations and constitutive relationships}

\subsection{Covariant decription of dielectric and
magnetic pro\-per\-ties of continuous media and vacuum}

The master equations of covariant electrodynamics of continuous  
media are the following \cite{landau,maug1}:
\begin{equation}
\nabla_{k} H^{ik} = - \frac{4\pi}{c} I^i \, ,
                                            \label{maxwell1}
\end{equation}
\begin{equation}
\nabla_{k} F^{*ik} =0 \,,
                                            \label{maxwell2}
\end{equation}
where $H^{ik}$ is the electric and magnetic induction tensor,
$\nabla_{k}$ is the covariant derivative, $I^i$ is the current
four-vector for free charges, $F^{*ik}$ is the tensor dual to the
Maxwell tensor $F_{mn}$.  The definition of duality for a generic
tensor $T^{ik}$ is
\begin{equation}
T^{*ik} = \frac{1}{2 \sqrt{-g}}\epsilon^{ikls} T_{ls},
                                            \label{dual}
\end{equation}
where $\frac{1}{\sqrt{-g}}\epsilon^{ikls}$ is the Levi-Civita tensor and
$\epsilon^{ikls}$ is the completely anti-symmetric symbol with
$\epsilon^{0123} = - \epsilon_{0123} = 1$.
As usually, we have
\begin{equation}
F_{ik} = \nabla_i A_k - \nabla_k A_i,
                                            \label{potential}
\end{equation}
where $A_i$ is the four-vector potential. With this definition   
equation (2) is trivially satisfied.

The set of equations (\ref{maxwell1}) and (\ref{maxwell2}) of
covariant electrodynamics of continuous media should be completed by
formulating consistently the constitutive equations 
\cite{landau,maug1,hehl1},
linking the electric and magnetic induction tensor with the Maxwell
tensor. The simplest relation between these tensors is linear and has
the following form:
\begin{equation}
H^{ik} = C^{ikmn}F_{mn}\, ,
                                            \label{constitutive1}
\end{equation}
where $C^{ikmn}$ is called the material tensor, describing the properties of
the linear response and containing the information about dielectric
and magnetic permeabilities, as well as about the magneto-electric
coefficients \cite{landau,maug1} (see, also, \cite{hehl1} for other details).
This tensor has the well-known index-transposition
properties:
\begin{equation}
C^{ikmn} = C^{mnik} = - C^{kimn} = - C^{iknm}.
                                            \label{symmetry1}
\end{equation}
Due to these symmetries the $C^{ikmn}$ tensor has 21 independent  
components.

The physical interpretation of the components of the material tensor
$C^{ikmn}$ is simplest in the frame of reference in which the medium
is at rest. Thus, the time-like four-vector $U^i$ of the medium 
is a necessary element of the theory.  When the medium is
present one has to include the $U^i$ vector into the constitutive
equations.  Note, however, that in vacuum there is no medium velocity
$U^i$ by definition, but one may still use the four-velocity of an
arbitrary observer.

The constitutive equations in the form (5) are phenomenological. This
phenomenological approach is appropriate to describe several types
of media, including the vacuum, which is usually considered as a sort
of medium in the framework of covariant electrodynamics of continuos
media.

\subsection{Decomposition of the material tensor  $C^{ikmn}$: 
the general case}

An interpretation of the 21 independent components of the material
tensor $C^{ikmn}$ can be given through relationships between the
four-vectors electric induction $D^i$ and magnetic field $H^i$, on one
hand, and the four-vectors electric field $E^i$ and the magnetic
induction $B^i$ on the other hand. In order to do this one defines the
vectors $D^i$, $H^i$, $E^i$ and $B^i$ by the following formulae
\cite{maug1}:
\begin{equation}
D^i = H^{ik} U_k, \quad
H^i = H^{*ik} U_k, \quad
E^i = F^{ik} U_k, \quad
B^i = F^{*ik} U_k.
                                            \label{constitutivevectors}
\end{equation}
These vectors are spacelike lying in a hypersurface orthogonal to
the velocity four-vector $U^i$,
\begin{equation}
D^i U_i = 0 = E^i U_i, \quad H^i U_i = 0 = B^i U_i \, .
                                            \label{orthogonality1}
\end{equation}
For $U^i$ timelike and normalized to unity, one can through 
formulae (7) and (8) represent the Maxwell tensor in the form
\begin{equation}
F_{mn} = \delta^{pq}_{mn}E_p U_q  - \epsilon_{mnls} \sqrt{-g} B^l U^s =
E_m U_n - E_n U_m - \eta_{mnl} B^l.
                                            \label{fdecomposition}
\end{equation}
where $\delta^{ik}_{mn}$ is the generalized 4-indices $\delta-$Kronecker 
tensor and $\eta_{mnl}$ is an anti-symmetric tensor orthogonal to $U^i$ 
defined as
\begin{equation}
\eta_{mnl} \equiv \sqrt{-g}\epsilon_{mnls} U^s, 
\quad
\eta^{ikl} \equiv \frac{1}{\sqrt{-g}}\epsilon^{ikls} U_s.
                                            \label{3levicivita}
\end{equation}
The generalized 6-indices $\delta-$Kronecker tensor and the spacelike
tensor $\eta^{ikl}$, both constructed from the Levi-Civita tensor, are
the standard tools for the decomposition of the material tensor
$C^{ikmn}$. They are connected by the useful identity
\begin{equation}
- \eta^{ikp} \eta_{mnp} = \delta^{ikl}_{mns} U_l U^s =
\Delta^i_m \Delta^k_n - \Delta^i_n \Delta^k_m ,
                                            \label{identity1}
\end{equation}
where the symmetrical projection tensor $\Delta^{ik}$ is defined as
\begin{equation}
\Delta^{ik} = g^{ik} - U^i U^k\, .
                                            \label{projector}
\end{equation}
It allows to project a generic tensor to a hypersurface orthogonal to
the four-velocity vector $U^i$.
Upon contraction, equation (11) yields another identity
\begin{equation}
\frac{1}{2} \eta^{ikl}  \eta_{klm} = - \delta^{il}_{ms} U_l U^s
= - \Delta^i_m .
                                            \label{identity2}
\end{equation}
We can now decompose $C^{ikmn}$ uniquely as
\begin{eqnarray}
&C^{ikmn} = \frac12 \left(
\varepsilon^{im} U^k U^n - \varepsilon^{in} U^k U^m +
\varepsilon^{kn} U^i U^m - \varepsilon^{km} U^i U^n \right) +
\nonumber \\&
+\frac12 \left[
-\eta^{ikl}(\mu^{-1})_{ls}  \eta^{mns} +
\eta^{ikl}(U^m\nu_{l \ \cdot}^{\ n} - U^n \nu_{l \ \cdot}^{\ m}) +
\eta^{lmn}(U^i \nu_{l \ \cdot}^{\ k} - U^k \nu_{l \ \cdot}^{\ i} )
\right] \, . &
                                            \label{maindecomposition}
\end{eqnarray}
Here $\varepsilon^{im}$, $(\mu^{-1})_{pq}$ and $\nu_{p \ \cdot}^{\ m}$ 
are defined as
\begin{equation}
\varepsilon^{im} = 2 C^{ikmn} U_k U_n, \quad
(\mu^{-1})_{pq} = - \frac{1}{2} \eta_{pik} C^{ikmn} \eta_{mnq}, 
                                           \label{epsilonmu}
\end{equation}
\begin{equation}
\nu_{p \ \cdot}^{\ m} = \eta_{pik} C^{ikmn} U_n 
=U_k C^{mkln} \eta_{lnp}
\, .
                                          \label{nu}
\end{equation}
The tensors $\varepsilon_{ik}$ and $(\mu^{-1})_{ik}$ are symmetric, but
$\nu_{l \ \cdot}^{\ k}$ is in general non-symmetric. The
dot denotes the position of the second index when lowered. 
These three tensors are spacelike, i.e., they are orthogonal to 
$U^i$, 
\begin{equation}
\varepsilon_{ik} U^k = 0, \quad (\mu^{-1})_{ik} U^k = 0, \quad
\nu_{l \ \cdot}^{\ k} U^l = 0 = \nu_{l \ \cdot}^{\ k} U_k.
                                           \label{orthogonality2}
\end{equation}
Calculating $D^i$ and $H_i$ given in (7), and using the decomposition  
(5) and (14), we obtain the standard linear relationships
\begin{equation}
D^i = \varepsilon^{im} E_m - B^l \nu_{l \ \cdot}^{\ i}, \quad
H_i = \nu_{i \ \cdot}^{\ m} E_m + (\mu^{-1})_{im} B^m \, .
                                           \label{dhlinearity}
\end{equation}
Thus, the tensors
$\varepsilon^{im}$,
$\mu_{pq}$ and
$\nu_{p \ \cdot}^{\ m}$
are the four-dimensional analogues of the dielectric permeability
tensor, the magnetic permeability tensor and the magneto-electric
coefficient tensor, respectively. One sees, that the 21 components of
$C^{ikmn}$ can be divided into the 6 components of $\varepsilon^{im}$,
the 6 components of $(\mu^{-1})_{pq}$ and the 9 components of $\nu_{p
\ \cdot}^{\ m}$. If the magneto-electric coefficient tensor 
$\nu_{i \ \cdot}^{\ m}$ is equal to zero, we have a special case of 
a so-called permeable medium and the $C^{iklm}$ tensor is characterized 
by 12 components only \cite{perlick}.

We now study four models where this decomposition takes place.

\subsection{Decomposition of the material tensor $C^{ikmn}$ in vacuum; its
dielectric properties in the presence of a gravita\-tional field}

\subsubsection{The first model: pure vacuum}

As a first model we consider the vacuum case in a generic curved
background.  One can ask what are the coefficients $C^{ikmn}$ for this
case.  For vacuum electrodynamics one has \cite{landau}
\begin{equation}
H^{ik} = F^{ik} = g^{im} g^{kn} F_{mn}.
                                           \label{h=f}
\end{equation}
Thus, from (\ref{constitutive1}), the material tensor $C^{ikmn}$ 
in vacuum must be the following
\begin{equation}
C_{(\rm vac)}^{ikmn} = \frac{1}{2} \left(g^{im} g^{kn} - g^{in}  
g^{km} \right).
                                           \label{cvacuum}
\end{equation}
We can see explicitly, that in vacuum the $U^i$ vector does not  
appear in $C^{ikmn}$.
From equations (\ref{epsilonmu})-(\ref{nu}) we get
\begin{equation}
\varepsilon^{ik} = (\mu^{-1})^{ik} = \Delta^{ik}, \quad
\nu_{i \ \cdot}^{\ k} = 0 \, .
                                           \label{epsilonmunuvacuum}
\end{equation}
From (\ref{epsilonmunuvacuum}) we see that the tensors
$\varepsilon^{ik}$ and $(\mu^{-1})^{ik}$ depend on the four-velocity,
in contrast to the $C_{\rm (vac)}^{ikmn}$ tensor (\ref{cvacuum}). Thus,
the vacuum in a gravitational field should be seen as a specific sort
of medium by specifying a four-velocity to it \cite{landau,mtw}.

\subsubsection{The second model: curvature induced corrections in  
vacuum}

As a second model we consider the one-loop corrections to quantum
electrodyna\-mics in a curved vacuum background \cite{drummond}. 
The relationship
between induction tensor and  Maxwell tensor has the form \cite{drummond}
\begin{equation}
H^{ik} = F^{ik}(1 + q_1 R) + q_2 (R^i_m F^{mk} - R^k_m F^{mi}) +
q_3 R^{ikmn} F_{mn},
                                           \label{hdrummond}
\end{equation}
where $R^{ikmn}$ is the Riemann tensor, $R^{ik}$ is the Ricci tensor,
$R$ is the scalar curvature of the spacetime, and the coefficients
\begin{equation}
q_1 = - \frac{\alpha \lambda^2_e}{180 \pi}, \quad
q_2 =  \frac{13 \alpha \lambda^2_e}{180 \pi}, \quad
q_1 = - \frac{\alpha \lambda^2_e}{90 \pi}
                                           \label{qcoefficients}
\end{equation}
contain the fine structure constant $\alpha$ and Compton wavelength of
the electron $\lambda_{e}$. From (\ref{hdrummond}) one sees that the
interaction between the electromagnetic field and curvature produces
an electric and magnetic polarization of the vacuum. Using
(\ref{constitutive1}), we can conclude that the material tensor is now
given by
\begin{equation}
C^{ikmn} =   C_{(\rm vac)}^{ikmn} +  C_{(\rm corr)}^{ikmn} \, ,
                                           \label{csum1}
\end{equation}
where, 
\begin{eqnarray}
& C_{(\rm corr)}^{ikmn} = 
\frac{1}{2} q_1 R \left(g^{im} g^{kn} - g^{in} g^{km} \right) +
\nonumber \\
&
+\frac{1}{2} q_2 \left( R^{im} g^{kn} - R^{in} g^{km} + R^{kn} g^{im} -
R^{km} g^{in} \right) + q_3 R^{ikmn} \, .&
                                           \label{ccorr1}
\end{eqnarray}
Note, that $C_{(\rm corr)}^{ikmn}$ is linear in the Riemann, Ricci
and scalar curvatures and does not contain the four-velocity 
vector $U^i$.

\subsection{Decomposition of $C^{ikmn}$  in a spatially  
isotropic medium}

\subsubsection{The third model: pure spatially isotropic medium 
in a gravitational field}

As a third model we consider a transparent spatially isotropic
medium.  From the phenomenological point of view a spatially isotropic
medium may be decribed by two scalars only: $\varepsilon$ - the
dielectric permeability and $\mu$ - the magnetic permeability.

Note the similarity of a spatially isotropic medium with its two 
scalars $\varepsilon$  and $\mu$, with the 
theory of elasticity also characterized by two scalars, 
the Lam\'e coefficients 
$\lambda$ and $\mu$. Thus, keeping this analogy, and recalling that 
the symmetric tensor of elastic modulus $E^{iklm}$ can be decomposed 
as $E^{iklm}=\lambda \Delta^{ik}\Delta^{lm} + \mu 
(\Delta^{il}\Delta^{km}+ \Delta^{im}\Delta^{kl})$, one can also 
decompose the anti-symmetric tensor $C^{iklm}$ using two tensorial 
quantities, $G^{ikmn}$ and $\Delta^{ikmn}$, defined by 
\begin{equation}
G^{ikmn} \equiv \frac{1}{2} \left(g^{im} g^{kn} - g^{in} g^{km}
\right)
                                           \label{glong}
\end{equation}
and $\Delta^{ikmn}$
\begin{equation}
\Delta^{ikmn} \equiv \frac{1}{2} \left(\Delta^{im} \Delta^{kn} -
\Delta^{in} \Delta^{km} \right)\,.
                                           \label{deltalong}
\end{equation}
These tensors form the basis for the decomposition of the 
anti-symmetric material tensor in a spatially
isotropic medium. We can then write the following linear combination
\begin{equation}
C^{ikmn}_{(\rm iso)} \equiv \frac{1}{\mu} \Delta^{ikmn} +
\varepsilon \left(G^{ikmn} - \Delta^{ikmn} \right) \, ,
                                           \label{cisotropic1}
\end{equation}
or in detailed form, 
\begin{eqnarray}
& C^{ikmn}_{(\rm iso)}=
\frac{1}{2\mu} \left(g^{im} g^{kn} -
g^{in} g^{km} \right) + \nonumber \\&
\left(\frac{\varepsilon \mu -1}{2\mu} \right)
\left(g^{im} U^k U^n - g^{in} U^k U^m + g^{kn} U^i U^m -
g^{km} U^i U^n \right) \, , &
                                           \label{cisotropic2}
\end{eqnarray}
where the coefficients in this linear combination were choosen 
to recover the standard definitions in three dimensions, i.e., 
\begin{equation}
D^i = \varepsilon E^i , \quad  H_i = \frac{1}{\mu} B_i , \quad
\varepsilon^{ik} = \varepsilon \Delta^{ik}, \quad
(\mu^{-1})^{ik} = \frac{1}{\mu} \Delta^{ik}, \quad
\nu_{i \ \cdot}^{\ k} = 0 .
                                           \label{dehb}
\end{equation}
If we put $\varepsilon$ and $\mu$ equal to unity, equation (\ref{cisotropic2}) 
yields,  the material tensor in vacuum (\ref{cvacuum}), as it should.

\subsubsection{The fourth model: curvature induced anisotropic
corrections in a spatially isotropic medium}

As before, through a phenomenological decomposition, we find the
anisotropic contributions to $C^{ikmn}$ from the Riemann tensor, 
the Ricci tensor and the scalar curvature. These contributions give 
\begin{equation}
C_{(\rm aniso)}^{ikmn} =   C_{(\rm iso)}^{ikmn} +  C_{(\rm corr)}^{ikmn} \, ,
                                           \label{csum2}
\end{equation}
where now,  
\begin{eqnarray}
&C^{ikmn}_{(\rm corr)} = \left( H_1 R + Q_1 R_{pq} U^p U^q \right)
\left(g^{im} g^{kn} - g^{in} g^{km} \right) +
\nonumber\\&
+\left( \bar{H}_1 R + \bar{Q}_1 R_{pq} U^p U^q \right)
\left(g^{im} U^k U^n - g^{in} U^k U^m + g^{kn} U^i U^m -
g^{km} U^i U^n \right)  +
\nonumber\\&
+Q_2 \left( R^{im} g^{kn} - R^{in} g^{km} + R^{kn} g^{im} - R^{km}  
g^{in} \right) +
\nonumber\\&
+\bar{Q}_2  \left(R^{im} U^k U^n - R^{in} U^k U^m + R^{kn} U^i U^m -
R^{km} U^i U^n \right) +
\nonumber\\&
+\hat{Q}_2 U^l \left( R^{(i}_l U^{m)} U^k U^n - R^{(i}_l U^{n)} U^k U^m +
R^{(k}_l U^{n)} U^i U^m - R^{(k}_l U^{m)} U^i U^n \right) +
\nonumber\\&
+Q_3 R^{ikmn} +
\nonumber\\&
+\bar{Q}_3 U_p U_q \left(R^{ipmq} g^{kn} - R^{ipnq} g^{km} +
R^{kpnq} g^{im} - R^{kpmq} g^{in} \right) +
\nonumber\\&
+\hat{Q}_3 U_p U_q \left(R^{ipmq} U^k U^n - R^{ipnq} U^k U^m +
R^{kpnq} U^i U^m - R^{kpmq} U^i U^n \right)\, ,
                                                 \label{ccorr2}
\end{eqnarray}
where, $(i \ m)$ denotes symmetrization in those indices, 
and the $H_1$ and $\bar{H}_1$, $Q_1, Q_2, Q_3$, $\bar{Q}_1, \bar{Q}_2,
\bar{Q}_3$ and $\hat{Q}_2, \hat{Q}_3$ coefficients are phenomenological 
parameters for the medium, which appear in this formalism in much the same 
way as $q_1, q_2, q_3$ have appeared in the vacuum case. However, 
we stress that contrary to the vacuum case where $q_1$, $q_2$, and $q_3$
can be found from within vacuum quantum electrodynamics \cite{drummond},
the coefficients  $H_1$, etc, cannot be directly extracted, since 
there is no developed self-consistent quantum electrodynamics 
in continuous media in curved spacetimes. 

The full material tensor (\ref{csum2}), $C_{(\rm aniso)}^{ikmn} =
C_{(\rm iso)}^{ikmn}+ C_{(\rm corr)}^{ikmn}$, containing
contribu\-tions from (\ref{cisotropic2}) and (\ref{ccorr2}), covers all
three previous examples (\ref{cvacuum}), (\ref{ccorr1}), and
(\ref{cisotropic2}), on choosing the coefficients in the decompositions
(\ref{cisotropic2}) and (\ref{ccorr2}) in an appropriate way. In this
sense it is the general construction to the four mentioned examples
given above.  For instance, for $\varepsilon = \mu = 1$ we may, if we
wish, restrict to the case where the coefficients $\bar{H}_1, Q_1,
\bar{Q}_1, \bar{Q}_2, \bar{Q}_3, \hat{Q}_2, \hat{Q}_3$ in
(\ref{ccorr2}) disappear. This could be achieved by puting these
coefficients proportional to $(\varepsilon \mu -1)^{\Gamma}$, ($\Gamma
> 0$).  Note, however, that from the phenomenological point of view
this requirement is not compulsory.

In order to be complete, we mention that the material tensor
$C^{iklm}$ we have been using can be enlarged to contain other
admissible non-minimal coupling terms \cite{hehl2}. Indeed, the most
general material tensor $\chi^{iklm}$, without curvature corrections,
has the generic form $\chi = f(x)\chi^0 + \alpha(x)[\epsilon]$, where
we have suppressed the indices, $[\epsilon]$ here means tensors
constructed from an odd number of the Levi-Civita tensor, and $x$
stands for spacetime coordinates. The scalar function $\alpha(x)$ is
an Abelian axion field which appears in some particle theories with
CP-violation \cite{hehl1,kim}.  Adding curvature corrections leads to
a non-minimal coupling contribution of duals of the curvature tensors,
$R^{*iklm}$ for instance, to the tensor $\chi^{iklm}$ \cite{hehl2}.
This approach would lead to an interaction term between the axions
and the curvature, but we do not investigate these models here.

\section{Exact solutions of Maxwell equations}

\subsection{Decomposition of the material tensor $C^{ikmn}$ for a
pp-wave gravitational background and its dielectric properties}

The previous sections contain the decomposition of the material tensor
for an arbitrary spacetime metric.  It is of interest to study the tensor
$C^{ikmn}$ in a particular background.  Here we will consider as the
spacetime background the exact  pp-wave solution of Einstein's
equations in vacuum described by the metric \cite{mtw,kram}
\begin{equation}
ds^{2} =  2 du dv -         
L^{2} \left[e^{2\beta}(dx^{2})^{2} + e^{-2\beta}(dx^{3})^{2} \right],
                                            \label{metric}
\end{equation}
where
\begin{equation}
u =  \frac{ct-x^{1}}{\sqrt{2}}, \quad  v = \frac{ct+x^{1}}{\sqrt{2}}
                                                 \label{time}
\end{equation}
are the retarded and the advanced time, respectively. The functions
$L$ and $\beta$ depend on $u$ only.

The pp-wave metric (\ref{metric}) possesses $G_5$ as the symmetry group 
\cite{kram} and admits the following set of Killing vector fields:
\begin{eqnarray}
&\xi^{i}_{(v)}= \delta ^{i}_{v}\,,  \quad
\xi^{i}_{(2)}= \delta ^{i}_{2}\,,  \quad
\xi^{i}_{(3)}= \delta ^{i}_{3}\,, \nonumber 
\\&
\xi^{i}_{(4)}= x^{2}  \delta^{i}_{v}  -  \delta^{i}_{2} 
\int{g^{22}(u)du}\,, \quad
\xi^{i}_{(5)} = x^{3}  \delta^{i}_{v}
-\delta^{i}_{3}  \int{g^{33}(u)du}\, ,
                                                 \label{killings}
\end{eqnarray}
of which the first three, $\xi^{i}_{(v)}$, $\xi^{i}_{(2)}$, and
$\xi^{i}_{(3)}$, form a $G_3$ Abelian subgroup of $G_5$.  Here
$g^{\alpha \beta}(u)$ ($\alpha, \beta = 2,3$) are the contravariant
components of the metric tensor. The vector $\xi^{i}_{(v)}$ is
isotropic, covariantly constant and orthogonal to the other four ones,
i.e., 
\begin{equation}
\nabla_{k} \  \xi^{i}_{(v)}=0\,, \quad 
g_{ik}  \xi^{i}_{(v)}  \xi^{k}_{(j)} =0 \, .
                                                 \label{orthogonality}
\end{equation}

The two functions  $L(u)$ and  $\beta(u)$  are coupled
by the equation:
\begin{equation}
L^{''} + L  (\beta^{'})^{2} = 0\,.
                                                \label{einstein}
\end{equation}
One can assume $\beta(u)$  as an arbitrary function of $u$ and 
then solve for $L$.
The curvature tensor has  two  non-zero  components:
\begin{equation}
- R^{2}_{\cdot u2u} = R^{3}_{\cdot u3u} =
L^{-2} \left[L^{2}  \beta^{'} \right]^{'} .
                                                \label{riemann}
\end{equation}
The Ricci tensor $R_{ik}$ and the curvature scalar $R$ are equal to zero.

For a medium (or an observer)  at rest in the chosen frame of  
reference one has
\begin{equation}
U^i = (\delta^i_u + \delta^i_v) \frac{1}{\sqrt{2}}\, .
                                                \label{velocity}
\end{equation}
Then one can show that for the pp-wave background, described by
equations (\ref{metric})-(\ref{riemann}), one obtains for a spatially
isotropic medium with curvature induced corrections (see
(\ref{csum2})-(\ref{ccorr2})), the following dielectric permeability,
magnetic permeability and magneto-electric coefficients, respectively:
\begin{eqnarray}
&\varepsilon^{im} = \varepsilon \Delta^{im} + 2 (Q_3 + \bar{Q}_3 +  
\hat{Q}_3)
R^{ikmn} U_k U_n\,, \nonumber
\\ &
(\mu^{-1})_{pq} = \frac{1}{\mu} \Delta_{pq} - \frac{1}{2}Q_3
\eta_{pik} R^{ikmn} \eta_{mnq} - 2 \bar{Q}_3 R_{plqs} U^l U^s \,,\nonumber
\\ &
\nu_{p \ \cdot}^{\ m} = Q_3 \ \eta_{pik} R^{ikmn} U_n\, .
                                                \label{gwepsilon}
\end{eqnarray}
Explicitly we find, 
\begin{eqnarray}
&
\varepsilon^{uu} = \varepsilon^{vv} = -
\varepsilon^{uv} = \frac{\varepsilon}{2}\,, \quad
(\mu^{-1})_{uu} = (\mu^{-1})_{vv} = - (\mu^{-1})_{uv} = \frac{1}{2\mu},
\nonumber
\\ &
\varepsilon^{u \alpha} = (\mu^{-1})_{u \alpha} =
\varepsilon^{v \alpha} = (\mu^{-1})_{v \alpha} = 0\,,  \quad
\varepsilon^{23} = (\mu^{-1})_{23} = 0 \,,
\nonumber
\\ &
\varepsilon^{22} =
g^{22} \left[ \varepsilon + (Q_3 + \bar{Q}_3 + \hat{Q}_3)  
R^2_{\cdot u2u} \right]\,, \quad 
\varepsilon^{33} =
g^{33} \left[ \varepsilon - (Q_3 + \bar{Q}_3 + \hat{Q}_3)  
R^2_{\cdot u2u}\right] \,, \nonumber
\\ &
(\mu^{-1})^{22} =
g^{22} \left[ \frac{1}{\mu} + (Q_3 - \bar{Q}_3) R^2_{\cdot u2u}  
\right]\,, \quad
(\mu^{-1})^{33} =
g^{33}\left[ \frac{1}{\mu} - (Q_3 - \bar{Q}_3) R^2_{\cdot u2u} \right] \,,
\nonumber
\\ &
\nu_{uu}=\nu_{uv}=\nu_{vu}=\nu_{vv}=\nu_{u \alpha}=\nu_{\alpha u}=
\nu_{v \alpha}=\nu_{\alpha v}=\nu_{22}=\nu_{33}= 0\,,
\nonumber
\\ &
\nu_{23} = \nu_{32} = - Q_3 L^2 R^2_{\cdot u2u}\,. &
                                                \label{explicit}
\end{eqnarray}
One sees, that the pp-wave field induces an anisotropy in the
dielectric properties of the initially spatially isotropic
medium. These anisotropies appear in the plane $x^2Ox^3$ only,
parallel to the gravitational wave front.

One important question arises: do these modifications in the
dielectric properties of the media produce changes in the
electromagnetic field structure and in its properties?  In order to
answer this question we consider now some exact solutions of Maxwell 
equations.

\subsection{States inheriting the symmetry of the GW background
and the reduction of Maxwell equations}

Searching for exact solutions of Maxwell equations 
(\ref{maxwell1})-(\ref{maxwell2}), it is advisable to fully 
use the inherited spacetime symmetries. Thus, we suggest first, 
that, for negative retarded time ($u\leq0$) before the appearance 
of the gravitational wave (at $u=0$)  the
dielectric medium is infinite, homogeneous and static. 
In other words, the tensor $C^{ikmn}$ can be put equal to a 
constant at $u \leq 0$.

For $u>0$, i.e., the GW already permeates the medium, the tensor
$C^{ikmn}$ contains the metric $g_{ik}(u)$ and the Riemann tensor
$R^{ikmn}(u)$ (see, e.g., (\ref{ccorr2})). Therefore, $C^{ikmn}$
depends on the retarded time only.  This idea can be formulated in a
covariant way.

Call ${\bf \Psi}$ an arbitrary macroscopic function of the state of
the electrodynamical system (material tensor, Maxwell tensor,
induction tensor, etc.), and, as before, let $\xi^i_{(a)}$ be the
Killing vectors belonging to the Abelian subgroup $G_3$ of the total
$G_5$ group of symmetry of the GW space-time. Now, if the
electrodynamical quantities are to inherit the spacetime symmetries
one must impose, ${\pounds}_{\xi_{(a)}} {\bf \Psi} =0$, where
${\pounds}_{\xi_{(a)}}$ stands for the Lie derivative along the
$\xi^i_{(a)}$ vector.  This zero Lie-derivative condition on the Maxwell and
induction tensors, i.e., ${\pounds}_{\xi_{(a)}}\ F_{mn} =0$ and
${\pounds}_{\xi_{(a)}}\ H_{mn} =0$, implies that they depend on the
variable $u$ only, $F_{mn}=F_{mn}(u)$ and $H_{mn}=H_{mn}(u)$.

Thus, the second subsystem of  Maxwell equations (\ref{maxwell2})
can be put in the form 
\begin{equation}
\frac{1}{2L^2} \frac{d}{du}\left(\epsilon^{iuls} F_{ls} \right) = 0\,.
                                          \label{transformedmaxwell}
\end{equation}
On integrating equation (\ref{transformedmaxwell}) we obtain, 
\begin{equation}
F_{v \alpha}(u) = F_{v \alpha}(0) = {\rm const}\,, \quad
F_{\alpha \beta}(u) =  F_{\alpha \beta}(0) = {\rm const}\,.
                                         \label{fieldconstant}
\end{equation}
Analogously, the first subsystem of Maxwell equations (\ref{maxwell1}) 
yields  
\begin{equation}
L^2 C^{iumn}(u) F_{mn}(u) =
C^{iumn}(0) F_{mn}(0) = {\rm const}.
                                         \label{inductionconstant}
\end{equation}
Only three equations of (\ref{inductionconstant}) are non-trivial (the
equations for $i=v,2,3$). The three functions $F_{u\alpha}$ and
$F_{uv}$ are unknown, since $F_{v \alpha}$ and $F_{\alpha \beta}$ are
constant (see (\ref{fieldconstant})) and can be expressed in terms of
initial data on the GW front.  Thus, we obtain the following reduced
algebraic system, containing three equations for the three unknown functions
\begin{eqnarray}
&C^{vuvu}(u) F_{uv}(u) + C^{vu \alpha u}(u) F_{u\alpha}(u) =
Z^v(u)\, , 
                                         \label{threesystem1}
\\
&
C^{\gamma uvu}(u) F_{uv}(u) + C^{\gamma u \alpha u}(u)
F_{u\alpha}(u) = Z^{\gamma}(u)\,, & 
                                         \label{threesystem2}
\end{eqnarray}
where
\begin{eqnarray}
&Z^v(u) = 
\frac{1}{L^2} C^{vuvu}(0) F_{uv}(0) +
\frac{1}{L^2} C^{vu \alpha u}(0) F_{u\alpha}(0) +\nonumber\\
&
+F_{v\alpha}(0) \left(C^{vuv\alpha}(u) -
\frac{1}{L^2} C^{vu v\alpha}(0) \right) + 
\frac{1}{2} F_{\alpha \beta}(0) \left(C^{vu\alpha \beta}(u) -
\frac{1}{L^2} C^{vu \alpha \beta}(0) \right)\,, &
                                         \label{zedv}
\end{eqnarray}
and,
\begin{eqnarray}
&Z^{\gamma}(u) = \frac{1}{L^2} C^{\gamma uvu}(0) F_{uv}(0) +
\frac{1}{L^2} C^{\gamma u \alpha u}(0) F_{u\alpha}(0) +\nonumber\\
&
+F_{v\alpha}(0) \left(C^{\gamma u v \alpha}(u) -
\frac{1}{L^2} C^{\gamma u v\alpha}(0) \right) + 
\frac{1}{2} F_{\alpha \beta}(0) \left(C^{\gamma u \alpha \beta}(u) -
\frac{1}{L^2} C^{\gamma u \alpha \beta}(0) \right)\,,
                                         \label{zedgamma}
\end{eqnarray}
are known functions of the retarded time $u$.

This system of three equations (\ref{threesystem1})-(\ref{threesystem2}) 
can be solved for generic material tensor (\ref{maindecomposition})
by the standard procedure involving Cramer's determinant, 
as in a linear algebra system of equations.

This system has an unique solution if the determinant ${\cal D}$ of
the $3\times3$ matrix $C^{iuku}(u)$ ($i,k=v,2,3$) is not identically
equal to zero.  For such a unique solution, the components $F_{uk}$ of
the Maxwell tensor contain the determinant ${\cal D}$ in their
denominators.

We introduce the term ``singular behaviour'' in the description of the
behaviour of the electromagnetic field, when the determinant ${\cal
D}$ is not equal to zero identically, but close to zero.  In
principle, using equation (\ref{threesystem1})-(\ref{threesystem2})
one can formulate explicitly this requirement in a general case for a
generic anisotropic medium.  Nevertheless, to clarify the physical
meaning of the singularities, we consider the four particular previous
examples of the $C^{iklm}$ tensor decomposition.

\subsection{Exact solutions in vacuum}

\subsubsection{The first model: pure vacuum}

The refractive index is defined by $n^2\equiv \varepsilon\mu$. For
pure vacuum $n^2=1$.  When $n^2= 1$, Cramer's determinant for
the system (\ref{threesystem1})-(\ref{threesystem2}) is equal to
zero. Then, equation (\ref{threesystem1}) gives
\begin{equation}
F_{uv}(u)=\frac{1}{L^2}F_{uv}(0)\, ,
                                \label{longelectric}
\end{equation}
and equation (\ref{threesystem2}) yields
\begin{equation}
F_{v\alpha}(0)=0 \, .
                                \label{transelectric}
\end{equation}
Due to the condition (\ref{transelectric}), the solution of the system
of equations (\ref{threesystem2}) exists if and only if
$F_{v\alpha}(u)=0$. On also has that $F_{23}$ and $F_{u\alpha}$ are
arbitrary functions of $u$, $F_{23}(u)$ and $F_{u\alpha}(u)$, since
they do not appear in the set of equations. This solution describes a
static longitudinal electric and magnetic field (longitudinal with
respect to the the direction of the GW propagation). However, the
transversal components of the field represent a travelling wave. Thus,
generally, when $n^2 = 1$, the required solution of Maxwell
equations, depending on the retarded time $u$ only, does not exist. In
other words, the problem of the response of the electrodynamic system
in vacuum (i.e., without dielectric material) is a subject of special
consideration, which we do now. 

Let us focus on the evolution of a magnetic field (we do not treat
here an electric field, since it yields some conceptual problems).
In the absence of GWs the magnetic field is static and homogeneous,
i.e., it means is constant throughout space.  Let us call the three
components of this constant field as $H^1$, $H^2$, and $H^3$.  In the
GW background metric (\ref{metric}) we have now to find a solution of
the Maxwell equations depending on the four coordinates
$(u,v,x^2,x^3)$, since there is no solution depending on the retarded
time $u$ alone.
In order to satisfy the initial conditions, i.e., in the absence of 
a GW the magnetic field is equal to an arbitrary constant vector, 
one can choose, for $u \leq 0$,  the components
of the electromagnetic potential four-vector $A_i$ depending 
linearly on the coordinates $(u,v,x^2,x^3)$:
\begin{equation}
A_u = A_v = 0, \quad A_2 = \frac{1}{\sqrt{2}} H^3 (u-v) + 
\frac{1}{2} H^1 x^3,
\quad
A_3 = \frac{1}{\sqrt{2}} H^2 (v-u) - \frac{1}{2} H^1 x^2.
                                         \label{initialdata}
\end{equation}
The Maxwell equations admit the solution in which $A_u$
and $A_v$ are not changed by the GW, and  the transversal components of the
four-vector of potential satisfy the equations \cite{balacqg}
\begin{equation}
\hat{D}A_{2} = 2 \beta^{'} \partial_{v} A_{2}\,, \quad
\hat{D}A_{3} = - 2\beta^{'} \partial_{v} A_{3}\,,
                                         \label{equationforpotential}
\end{equation}
where $\hat{D} \equiv 2\partial_{u}\partial_{v} +
g^{\alpha\beta}\partial_{\alpha}\partial_{\beta}$.
It is easy to check that the formulae
\begin{eqnarray}
&F_{uv} \equiv F_{01} = 0,\quad
F_{32} = H^1 , \quad
\nonumber\\&
F_{v2} = - \frac{1}{\sqrt{2}} H^3 e^{\beta},\quad
F_{v3} = \frac{1}{\sqrt{2}} H^2 e^{- \beta},
\nonumber\\&
F_{u2} = - F_{v2} + \beta^{'} e^{\beta} M_2 , \quad
F_{u3} = - F_{v3} - \beta^{'} e^{- \beta} M_3 \,,&
                                         \label{vacuumsolution1}
\end{eqnarray}
where $M_2$ and $M_3$ are defined by
\begin{eqnarray}
&M_2 \equiv \frac{1}{\sqrt{2}} H^3 (u-v) + \frac{1}{2} H^1 x^3 e^{- \beta}\,, 
\nonumber\\&
M_3 \equiv \frac{1}{\sqrt{2}} H^2 (v-u) - \frac{1}{2} H^1 x^2 e^{\beta}\,,&
                                         \label{vacuumsolution2}
\end{eqnarray}
represent the solution of equations (\ref{maxwell1})-(\ref{maxwell2}) 
in vacuum with initial conditions (\ref{initialdata}).

Formulae (\ref{vacuumsolution1})-(\ref{vacuumsolution2}) 
recover the known results of \cite{boc,zel1,zel2},
when we reduce  the nonlinear GW background to the case of a weak GW. It  
is important to emphasize that the gravitationally induced electric  
field, whose components are, 
\begin{equation}
E_2 = -\frac{1}{\sqrt{2}} \beta^{'} e^{\beta} M_2 \,, \quad
E_3 = \frac{1}{\sqrt{2}} \beta^{'} e^{- \beta} M_3 \,,
                                         \label{vacuumsolution3}
\end{equation}
grows linearly as a function of the coordinates $(v,x^2,x^3)$.

\subsubsection{The second model: curvature induced corrections in  
vacuum}

From equations (\ref{threesystem1}) one obtains equation
(\ref{longelectric}), as in the previous case.  Then equation
(\ref{threesystem2}) also coincides with equation
(\ref{transelectric}) for pure vacuum because the curvature induced
corrections for the material tensor (see equation (\ref{ccorr1})) do
not give any contribution to $C^{iuku}$ tensor in equations
(\ref{threesystem1})-(\ref{threesystem2}). As in the previous case
there is no solution depending on $u$ only. For a different solution
see \cite{balacqg}.

\subsection{Exact solutions in a spatially isotropic medium}

\subsubsection{The third model: pure spatially isotropic medium 
in a gravitational wave field}

Consider now a spatially isotropic medium with $n^2\neq1$.  Using
formula (\ref{cisotropic2}) for the $ C^{ikmn}$ coefficients and
equation (\ref{velocity}) for the four-velocity of the medium, one can
extract that equation (\ref{threesystem1}) also gives equation
(\ref{longelectric}). The remaining pair of equations
(\ref{threesystem2}) takes the form
\begin{equation}
(n^2-1) g^{\gamma\alpha} F_{u\alpha}(u) = (n^2 -1)
\frac{\eta^{\gamma\alpha}}{L^2} F_{u\alpha}(0) +
(n^2+1) F_{v\alpha}(0)
\left(g^{\gamma\alpha} - \frac{1}{L^2} \eta^{\gamma\alpha} \right)\, ,
                                     \label{transeq1}
\end{equation}
where $\eta_{ik}$ is the Minkowski metric.

For $n^2 \neq 1$, the solution of the system (\ref{transeq1}) 
is unique and equal to 
\begin{eqnarray}
&F_{u2}(u) = F_{u2}(0) + F_{u2}(0)\left( e^{2\beta} - 1 \right) +
\left(\frac{n^2+1}{n^2-1}\right)
F_{v2}(0)\left( e^{2\beta} - 1 \right)\,, 
\nonumber\\&
F_{u3}(u) = F_{u3}(0) + F_{u3}(0)\left( e^{-2\beta} - 1 \right) +
\left(\frac{n^2+1}{n^2-1}\right)
F_{v3}(0)\left(e^{-2\beta} - 1 \right)\,. &
                                       \label{transfield1}
\end{eqnarray}
The physical analysis of the solution 
(\ref{longelectric}) and (\ref{transfield1}) is more transparent  
in terms of the electric and magnetic four-vectors 
(\ref{constitutivevectors}):
\begin{eqnarray}
&E_1(u) \equiv F_{10} \equiv -F_{uv} = \frac{1}{L^2} E_1(0)\,,
\nonumber\\&
B^1(u) \equiv F_{32} = B^1(0),
\nonumber\\&
E_2(u) \equiv F_{20} \equiv \frac{1}{\sqrt{2}}(F_{2u} + F_{2v}) =
E_2(0) + \frac{1}{(n^2-1)} \Pi_2(u),
\nonumber\\&
B^2(u) \equiv F_{13} \equiv \frac{1}{\sqrt{2}}(F_{v3} - F_{u3}) =
B^2(0) + \frac{1}{(n^2-1)} \Pi_3(u),
\nonumber\\&
E_3(u) \equiv F_{30} \equiv \frac{1}{\sqrt{2}}(F_{3u} + F_{3v}) =
E_3(0) + \frac{1}{(n^2-1)} \Pi_3(u),
\nonumber\\&
B^3(u) \equiv F_{21} \equiv \frac{1}{\sqrt{2}}(F_{u2} - F_{v2}) =
B^3(0) - \frac{1}{(n^2-1)} \Pi_2(u)\,, &
                                                    \label{eb}
\end{eqnarray}
where
\begin{eqnarray}
&\Pi_2(u) = \left(e^{2\beta}-1 \right)
\left[n^2 E_2(0) + B^{3}(0) \right],
\nonumber\\&
\Pi_3(u) = \left(e^{-2\beta}-1 \right)
\left[n^2 E_3(0) - B^{2}(0) \right] \, .
                                                    \label{pi2pi3}
\end{eqnarray}
We see that when $n^2 \to 1$, then the gravitationally induced part of
the electric and magnetic fields, i.e., the terms involving
$\frac{1}{(n^2-1)} \Pi_\alpha(u)$, go to infinity (we are not taking
$n^2 E_2(0) = - B^3(0)$, and $n^2 E_3(0) =
B^2(0)$ simultaneously, which is a rather special case, of no interest
here). Thus, we have two distinct situations (i) if we take first the
limit of $\beta \to 0$ and then $n^2\to 1$ we obtain that the
gravitational induced part of the Maxwell tensor $F_{ik}$ is zero,
i.e., the double limit: 
$\lim_{n^2 \to 1} \lim_{\beta \to 0} \{F_{ik}\}=0$, where 
$\{F_{ik}\}$ denotes the gravitational induced
part of the Maxwell tensor $F_{ik}$; (ii) if we take first the limit
of $n^2\to 1$ and then $\beta \to 0$  we obtain that the
gravitational induced part of the Maxwell tensor $F_{ik}$ is 
infinite, i.e., the double limit
$ \lim_{\beta \to 0} \lim_{n^2 \to 1} \{F_{ik}\}=\infty$.
Since these limits do not coincide, we can speak of a critical  
behaviour of the electromagnetic field near the singular point  
$n^2=1$. In the absence of the GW such a problem does not arise.

\subsubsection{The fourth model: curvature induced anisotropic
corrections in a spatially isotropic medium}

When the tensor of linear response $C^{ikmn}$ 
is given by the formulae (\ref{csum2})-(\ref{ccorr2}), 
then the longitudinal component of the electric field  
coincides with (\ref{longelectric}), 
but the remaining two-component subsystem of  
equations (\ref{threesystem2}) takes the form
\begin{eqnarray}
&\left[ (n^2-1) g^{\gamma\alpha} + \mu \hat{Q}_3 R^{\gamma v \alpha  
v}\right]
F_{u\alpha}(u) = 
(n^2+1) F_{v\alpha}(0) \left(g^{\gamma\alpha} -
\frac{1}{L^2} \eta^{\gamma\alpha} \right) +
\nonumber\\&
+(n^2-1) \frac{\eta^{\gamma\alpha}}{L^2} F_{u\alpha}(0) 
- \mu F_{v\alpha}(0)
R^{\gamma v \alpha v}(\hat{Q}_3 + 2\bar{Q}_3) \,. &
                                           \label{transeq2}
\end{eqnarray}
In this case Cramer's determinant ${\cal D}$ of the system 
(\ref{transeq2}) is equal to
\begin{equation}
{\cal D} = \frac{1}{L^4}
\left[(n^2 - 1) + \mu \hat{Q}_3 R^2_{\cdot u2u} \right]
\left[(n^2 - 1) + \mu \hat{Q}_3 R^3_{\cdot u3u} \right].
                                          \label{determinant1}
\end{equation}
Recalling that $R^2_{\cdot u2u} = - R^3_{\cdot u3u}$, we  
find 
\begin{equation}
{\cal D} = \frac{1}{L^4}
\left[
(n^2 - 1)^2 -  \mu^2 \hat{Q}^2_3 (R^2_{\cdot u2u})^2 \right]\,,
\end{equation}
and
\begin{eqnarray}
&
F_{u2}(u) = \left[1 + \frac{\mu \hat{Q}_3}{(n^2-1)} R^{2}_{\cdot  
u2u} \right]^{-1}
\left\{
F_{v2}(0) \frac{1}{(n^2-1)} \left[(n^2+1)\left(1 - e^{2\beta} \right) 
\right. \right.
\nonumber\\&
\left. \left.
- \mu R^{2}_{\cdot u2u}(\hat{Q}_3 + 2\bar{Q}_3) \right] +
F_{u2}(0) e^{2\beta} \right\} \,,
\nonumber\\&
F_{u3}(u) = \left[1 + \frac{\mu \hat{Q}_3}{(n^2-1)} R^{3}_{\cdot  
u3u} \right]^{-1}
\left\{ 
F_{v3}(0) \frac{1}{(n^2-1)} \left[(n^2+1)\left(1 - e^{-2\beta} \right) 
\right. \right.
\nonumber\\&
\left. \left.
-\mu R^{3}_{\cdot u3u}( \hat{Q}_3 + 2\bar{Q}_2) \right] +
F_{u3}(0) e^{-2\beta}  \right\}\,. &
\end{eqnarray}
There is a singularity when ${\cal D} \to 0$. For instance, this can
happen, when $n^2$ is near one and the quantity $\mu^2 \hat{Q}^2_3
(R^2_{\cdot u2u})^2$ is equal to the small difference $(n^2-1)^2$.
Since the Riemann tensor can periodically change its value, due to the
periodic oscillations of the GW, one could observe a periodic
increasing of one of the two component of the electric and magnetic
fields. This type of singularities is very similar to the behaviour of
the previous third model.  The novelty here, is that periodic
variations of the curvature tensor produce a natural approach to the
critical point at ${\cal D} =0$, which repeats itself periodically.
In principle, from the phenomenological point of view, we can 
consider a new sort of medium with $n^2=1$, $\mu=1$, and
$\hat{Q}_3\neq0$. Then ${\cal D}=-\hat{Q}^2_3
(R^2_{\cdot u2u})^2/L^4$. From equation (\ref{ccorr2}), we see that for 
$\hat{Q}_3\neq0$ the material tensor depends on the velocity $U^i$,
which is a characteristic of media in general. Therefore, we 
can call this medium a ``quasi-vacuum''.

\section{Conclusions}

We have shown that there exists a new class of solutions  
of evolutionary equations for initially static electric and  
magnetic fields inside a media in a GW background. 
The solutions on this class have two important properties: the first
property is that they inherit the symmetry of the GW field and depend
on the retarded time only. In vacuum there are no such solutions, as we
have shown. The second property is connected with the singularities
which appear in the electro-magnetic response to the GW. These
singularities have a general status and appear not only in a material
media with arbitrary spatial symmetry, but also in a ``quasi-media'',
such as vacuum interacting with curvature. 
This second property of the solutions gives a clear motivation to  
search for experimental confirmation of this theoretical  
predictions, namely the amplification of  electric and 
magnetic fields due to the passage of a GW.

\section{Acknowledgments}
The authors are grateful to Richard Kerner for fruitful discussions. 
AB thanks the hospitality  CENTRA/IST in Lisbon, and a special 
grant from ESO/FCT to invited scientists. 
This work was partially funded by FCT through project ESO/PRO/1250/98.


\begin{thebibliography}{50}

\bibitem{boc}  D. Bocaletti, V. De Sabbata, P. Fortini, C. Gualdi, 
\textbf{B70} (1970) 129.

\bibitem{zel1} Ya. B. Zel'dovich, Sov. Phys. JETP \textbf{38} (1974) 652.



\bibitem{zel2} Ya. B. Zel'dovich, I. D. Novikov, 
{\it The structure and evolution of  the Universe}
(Chicago University Press, Chicago 1983), p.472.

\bibitem{grishpoln} L. P. Grishchuk, A. G. Polnarev, in {\it General
Relativity and Gravitation, One Hundred Years After the Birth of
Albert Einstein} Vol. II,  ed. A. Held, (Plenum Press, New York,
1980), p. 393.

\bibitem{bala} A. B. Balakin,  D. V. Vakhrushev,  
Russian Physics Journal \textbf{36} (1993) 833.

\bibitem{landau} L. D. Landau, E. M. Lifchitz, L. P. Pitaevskii,  
{\it Electrodynamics of Continuous Media} 
(Butterworth Heinemann, Oxford 1996).

\bibitem{maug1} G. A. Maugin, J. Math. Phys. \textbf{19} (1978) 1198.

\bibitem{hehl1} F. W. Hehl, Yu. N. Obukhov, G. F. Rubilar, 
Ann. Phys. (Leipzig) \textbf{11} (2000) Spec. Issue 71.

\bibitem{perlick} V. Perlick, {\it Ray Optics, Fermat's Principle 
and Applications to General Relativity} (Springer-Verlag, Berlin 2000). 

\bibitem{mtw} C. W. Misner, K. S. Thorne, J. A. Wheeler, {\it Gravitation},
(Freeman, San Francisco 1973).

\bibitem{drummond} I. T. Drummond, S. J. Hathrell,  Phys. Rev. D \textbf{22} 
(1980) 343. 

\bibitem{hehl2} F. W. Hehl, Yu. N. Obukhov, in {\it 
Testing Relativistic Gravity in Space}, eds. C. Laemmerzahl et al 
(Springer-Verlag 2000). 

\bibitem{kim} J. E. Kim,  ``Axion theory review'', astro-ph/9812257.

\bibitem{kram} D. Kramer, H. Stephani, M. McCallum, E. Herlt, E. Schmutzer, 
{\it Exact solutions of the Einstein field   equations}
(Cambridge University Press, Cambridge 1980). 

\bibitem{balacqg} A. B. Balakin,  Class. Quantum Grav. 
\textbf{14} (1997) 2881.

\end{thebibliography}
\end{document}